\documentclass[12pt,preprint]{aastex}
\usepackage{emulateapj5}
\begin {document}

\title{Near-Contemporaneous Optical Spectroscopic and Infrared Photometric Observations of Candidate Herbig Ae/Be Stars in the Magellanic Clouds}

\author{Bradley W. Rush\altaffilmark{1}, John P. Wisniewski\altaffilmark{2,3,4}, Karen S. Bjorkman\altaffilmark{1,3}}

\altaffiltext{1}{Dept of Physics \& Astronomy, University of Toledo, 2801 W. Bancroft St., Toledo, OH 43606-3390; Bradley.Rush@rockets.utoledo.edu, Karen.Bjorkman@utoledo.edu}
\altaffiltext{2}{Department of Astronomy, University of Washington, Box 351580, Seattle, WA 98195; jwisnie@u.washington.edu}
\altaffiltext{3}{Visiting Astronomer, Cerro Tololo Inter-American Observatory, National Optical Astronomy Observatory, which is operated by the Association of Universities for Research in Astronomy, under contract with the National Science Foundation.}
\altaffiltext{4}{NSF Astronomy \& Astrophysics Postdoctoral Fellow}

\begin{abstract}

We present near-IR (J,H,Ks) photometry for 27 of the 28 candidate Herbig Ae/Be stars in the Small and Large Magellanic Clouds identified via the 
EROS1 and EROS2 surveys as well as near-contemporaneous optical (H$\alpha$) spectroscopy for 21 of these 28 candidates.  Our observations extend 
previous efforts to determine the evolutionary status of these objects.  We compare the IR brightness and colors of a subset of our sample 
with archival ground-based IR data and find evidence of statistically significant photometric differences for ELHC 5, 7, 12, 18, and 21 in one or more filter.  
In all cases, these near-IR photometric variations exhibit a grey color as compared to earlier epoch data.  The $\sim$1 magnitude IR brightening and minimal change in the H$\alpha$ emission strength we observe in ELHC 7 is consistent with previous claims that it is a UX Ori type HAe/Be star, which is occasionally obscurred by dust clouds.  We also detect a $\sim$1 magnitude IR brightening of ELHC 12, but find little evidence of a similar large-scale change in its H$\alpha$ line strength, suggesting that its behavior could also be caused by a UX Ori-like event.  The $\sim$0.5 magnitude IR variability we observe for ELHC 21, which also exhibited little evidence of a change in its H$\alpha$ emission strength, could conceivably be caused by a major recent enhancement in the density of the inner disk region of a classical Be star.  We also report the first near-IR photometry for two ESHC stars and the first H$\alpha$ spectroscopy for one ELHC and five ESHC stars.  Although H$\alpha$ emission is detected in all of these new observations, they do not exhibit a strong near-IR excess.  It is therefore possible that many of these objects may be classical Be stars rather than Herbig Ae/Be stars.

\end{abstract}


\section{Introduction}

Herbig Ae/Be stars \citep{her60}, or HAeBe stars, are intermediate mass pre-main-sequence A or B type stars, analogous to low-mass T Tauri stars.  HAeBe stars exhibit photometric, spectroscopic, and polarimetric variability, may be co-located with active star formation regions or be isolated, are luminosity class III to V, and exhibit an infrared (IR) excess due to thermal emission from circumstellar dust \citep{gri94,wat98,oud01}.  Some of these observational characteristics are also shared with objects of different evolutionary status such as classical Be stars (see e.g. \citealt{por03}).  Thus distinguishing between HAeBe and classical Be stars can be a challenging observational endeavor \citep{bea01,dew05}.

Metallicity is believed to influence many aspects of stellar astrophysics, ranging from the expected prevalence of the formation of extrasolar planets
\citep{gon97,fis05} to the basic evolution of stars \citep{mey94,chi06}.  For circumstellar disk systems, the lower metallicity of the Small and Large Magellanic 
Clouds (SMC and LMC) affects the density and/or temperature of gaseous decretion disks as compared to our Galaxy \citep{wis07}.  For dusty circumstellar 
disk systems, metallicity will also affect observables such as the magnitude of the IR excess of each system (see e.g. \citealt{dew03} and \citealt{dew05}) and 
may result in a higher luminosity compared to their Galactic counterparts \citep{bea01}.

While large-scale IR surveys of our Galaxy have revealed a large population of intermediate- to high-mass young stellar objects \citep{rob08,urq10}, the analogous population of HAeBe stars in the metal poor environments of the Small and Large Magellanic Clouds (SMC and LMC) have yet to be identified in a complete and systematic manner, thus hampering efforts to observationally constrain the process of star and planet formation in low metallicity environments.  A growing number of candidate SMC and LMC intermediate mass young stars were identified via the EROS1 and EROS2 surveys \citep{lam99,bea01,dew02,dew03} as well as surveys using the Spitzer space telescope \citep{whi08,cla10}.  However follow-up observations of many of the less embedded HAeBe candidates have raised questions about whether the objects are HAeBe stars or classical Be stars \citep{dew03,dew05,wis06,bjo05}, thereby motivating additional investigations.

Accurately characterizing the IR properties of candidate HAeBe stars is one means to better constrain their classification, as thermal emission from dust should exhibit a large near/far IR excess, whereas ``contaminant'' objects such as classical Be stars should exhibit only a moderate near-IR excess due to bound-free/free-free radiation.  Both the limited mid-IR (e.g. 24 $\mu$m) sensitivity and angular resolution ($\sim$2$\farcs$0 in IRAC bands; $\sim$6$\farcs$0 in the 24$\mu$m MIPS band) of surveys such as the Spitzer SAGE survey \citep{mei06} have precluded stringent constraints being placed on many SMC and LMC candidate HAeBe stars identified previously via the EROS survey \citep{lam99,dew02}, although several candidates are in the SAGE catalog \citep{whi08}.  Interestingly, $\sim$3\% of the SAGE young stellar object candidate list exhibits variability in 2-epoch observations at Spitzer wavelengths \citep{vij09}.

In this paper, we present ground-based near-IR (JHK) photometry of 27 of the 28 candidate HAeBe stars identified in the EROS survey with both high photometric accuracy and at small angular resolution.  These data allow us to explore source confusion in earlier works and identify and characterize IR variability.  The dataset also provides a high quality baseline of near-IR data for future detailed studies of individual candidates.  We supplement these results with new, near-contemporaneous H$\alpha$ spectroscopic observations for 21 of the 28 candidates.  In Section 2 we outline our observations and data reduction processes.  We describe our results on source confusion and variability in Section 3, and discuss our overall results in Section 4.

\section{Observations and Data Reduction}  \label{obsection}
\subsection{IR Photometry}

We observed 27 of the 28 EROS SMC HAeBe Candidates (ESHC) and EROS LMC HAeBe candidates (ELHC) \citep{lam99,bea01,dew02,dew03} in the J (1.25 $\mu$m), H (1.635 $\mu$m), and Ks (2.150 $\mu$m) bands.  The observations were made on 2005 January 9, using the Infrared Side Port Imager (ISPI) instrument \citep{van04} on the Cerro Tololo Inter-American Observatory (CTIO) 4-m telescope (Table \ref{targetcoords}).  The data were recorded on a Hawaii-2 array with a pixel scale of 0.3 arcsec pixel$^{-1}$, yielding a field of view of 10$\farcs$25 x 10$\farcs$25 arcmin.  The typical seeing during the night varied between 0$\farcs$6 and 1$\farcs$2.  Observations of our standard star and science fields were obtained using a Fowler sample of 1, using the number of coadds and integration per coadd listed in Table \ref{zps}.  We dithered the telescope after each integration to facilitate more accurate sky subtraction.  Flat fielding was achieved by subtracting a series of dome flats with an illuminating lamp turned off from a series with the lamp turned on.

These data were reduced using \textit{cirred}, a custom CTIO IR reduction package for IRAF\footnote{IRAF is distributed by the National Optical Astronomy Observatory, which is operated by the Association of Universities for Research In Astronomy (AURA) under cooperative agreement with the National Science Foundation}.  Our data were first divided by average flat field images, constructed via the method described above.  To perform sky subtraction, we computed the average background count of each image with all stars masked, and then replaced the pixel value of each star with the image's average background.  All images in a dither set were then averaged to create a master sky-subtraction image for the group, and this master sky image was then subtracted from each image.  A World Coordinate System (WCS) was assigned to each image using SExtractor \citep{ber96} to identify stellar centroid positions and WCSTools \citep{min97} to correlate these with the $2MASS$ database.  The high order distortions of ISPI were computed and corrected using WCSTools and the IRAF routine \textit{ccmap}.  Finally, dithered images were registered and averaged using the program SWARP \citep{ber02}.

We first performed aperture photometry on all stars in our data using a 3$\farcs$6 (12 pixel) radius aperture and an annulus of radial size 3$\farcs$6-6$\farcs$0 (12-20 pixels) for background count determination.  Photometric calibration to an absolute scale was achieved by comparing $\sim$20 stars per field of view, distributed across the entire array, to published 2MASS photometry \citep{skr06}.  The average difference between the instrumental photometry and archival 2MASS photometry of these stars defined the zero point for each of our fields of view (Table \ref{zps}), and the standard deviation of this average defined the respective error in each zero point (Table \ref{zps}).  The accuracy of this method was checked by comparing our observations of select near-IR photometric standards from the \citet{per98} catalog to published 2MASS values, and as seen in Table \ref{stds} these results are consistent to within the error of the measurements.  To mitigate source confusion problems, we next computed an aperture correction to a 1$\farcs$2 (4 pixel) radius aperture and extracted aperture photometry for all sources using this smaller aperture.  The photometry reported in this paper utilizes this smaller, 1$\farcs$2 aperture size.

\subsection{Optical Spectroscopy}
We observed 21 of the 28 ESHC and ELHC stars on 2005 January 21 using Hydra, the 138 fiber multi-object spectrograph on the 4-m telescope at CTIO.  The observations, summarized in Table \ref{spec} were made with the KPGL3 grating and 200 $\mu$m slit plate, providing coverage from $\sim$3900 - 6700 \AA\ at R$\sim$2200.  The width of the science fibers was 2$\farcs$0 in diameter.  10-15 blank-sky fibers were assigned per fiber configuration to measure the nebular sky emission near our targets.  We used 3600 second integration times, yielding the signal to noise (SNR) per pixel, measured in a line free continuum normalized region near 6500\AA, shown in Table \ref{spec}.  Calibration 
data including bias frames, dome flats, and HeNeArXe ``penray'' lamp exposures were obtained, along with sky flats to measure fiber-to-fiber throughput, projector flats to trace the location of each fiber orientation, and milk flats to create bad pixel masks.  

The data were reduced in IRAF using the \textit{dohydra} package and followed standard practices for multi-object spectroscopy.  Sky subtraction was performed in an interactive manner, using different nearby sky fibers to best subtract the nebular emission in each science fiber.  Our choice of continuum placement is the dominant source of error in our quoted equivalent width (EW) in Table \ref{spec}; we estimate this uncertainty to be $<$5\% by computing EWs for a number of sources using a range of continuum definitions.

\section{Results}

\begin{figure}
\begin{center}
\includegraphics[scale=0.5]{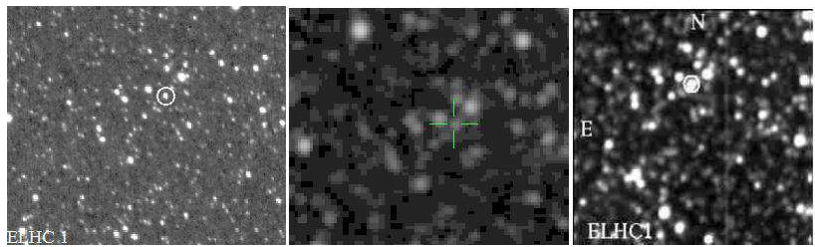}
\caption{The finder chart for ELHC 1 from our J-band ISPI data (left panel), J-band 2MASS data (middle panel), and optical R$_{E}$ band imagery (right panel) from the EROS survey is shown.  The EROS survey finder chart is reproduced with permission from de Wit et al (2002), A\&A, 395, 829 (copyright ESO).  The field of view in each panel is 1$^{'}$5 x 1$^{'}$5.  Finder charts for other targets are available in the online version of this manuscript. \label{finder}}
\end{center}
\end{figure}

Given the pixel size of the CCD detectors in the EROS1 (1$\farcs$25; \citealt{lam99}) and EROS2 (0$\farcs$6; \citealt{pal98}) and typical seeing of 2$\farcs$0 in these surveys \citep{pal98}, source confusion for the ESHC and ELHC targets can be a concern.  After carefully cross-correlating our imagery with published finder charts (e.g. \citealt{dew02}), we extracted photometry for 27 of 28 ESHC/ELHC targets (see Table 5).  Note that our photon statistic errors were typically of order $\sim$0.005 mag, which is an order of magnitude less than the uncertainty in our zero points (Section \ref{obsection}); hence we simply adopted the zero point uncertainty as the errors in our absolute photometry (Table \ref{phot}).  Updated coordinates, accurate to 0.1 seconds in RA and 0$\farcs$4 in Dec, for each candidate are provided in Table \ref{targetcoords}, while updated J-band finder charts are provided in Figure \ref{finder}.

\subsection{Source Confusion} \label{confuse}

Potential source confusion has been noted in some follow-up investigations of ELHC stars, with \citet{dew05} reporting concerns with ELHC 5, 6, 8, and 18 (the latter of which exhibited 2 images in their near-IR imagery) and \citet{wis06} reporting that ELHC 11 might have two components.  The updated coordinates (Table \ref{targetcoords}) and finder charts (Figure 1) provided by our better resolution data should help alleviate source identification problems for future investigations.

We found clear evidence in our data that ELHC 6 exhibited two sources at the position reported by \citet{dew02} (see Table \ref{phot}), with brightnesses corresponding to 
15.66$\pm$0.03 and 16.10$\pm$0.03 at J, 15.42$\pm$0.03 and 15.63$\pm$0.03 at H, and 15.42$\pm$0.03 and 15.86$\pm$0.03 at Ks.  The PSF shape of ELHC 18 was also elongated and likely comprised of two sources, confirming the confusion reported by \citet{dew05}.  ELHC 4 and ELHC 11 also had clear evidence of nearby companions at radial distances of $\sim$1$\farcs$2 and $\sim$1$\farcs$5 away respectively.  Careful examination of ELHC 5 and ELHC 8, reported to have potential nearby neighbors \citep{dew05}, revealed no evidence of source contamination in our data.

\subsection{IR Photometric Variability}

IR photometric variability has been previously observed in both Galactic \citep{eir02,bar09,muz09} and Magellanic Cloud \citep{vij09} pre-main sequence disk systems.  While ESHC and ELHC stars were initially identified as potential pre-main sequence objects based on their optical photometric variability, the stability of their IR brightness has not been well constrained to date.  We limit our investigation of IR photometric variability to sources which have archival ground-based near-IR observations (e.g. \citealt{dew05}) with adequate spatial resolution to mitigate the source confusion that likely affects at least some of the quoted 2MASS measurements of these stars.  Thus, we do not use our IR photometry of ESHC sources to identify variability in these objects.

We subtracted the archival photometry for 15 ELHC stars from our measurements given in Table \ref{phot}, and compile the results in Table \ref{photdiff}.  Although most of the ELHC stars exhibit stable absolute photometry at the level of our uncertainties, ELHC 3, 5, 7, 12, 18, and 21 exhibit 
photometric variability at the $>$3$\sigma$ level in at least one filter.  Three sources exhibit strong variability in all 3 IR filters, ELHC 21 ($\sim$0.5 magnitude difference), ELHC 7 ($\sim$1 magnitude diference), and ELHC 12 ($\sim$1 magnitude difference).  ELHC 18 was reported to have clear source confusion issues in Section \ref{confuse}; hence, we limit our interpretation of the origin of variability originating from this set of objects.  We do note a hint of a trend in our photometry as compared to \citet{dew05} in Table 
\ref{photdiff}, such that we tend to report more ELHCs to be brighter (11 of 15) than fainter (4 of 15) as compared to the archival values.  Our standard star observations do match extremely well with published values (Table \ref{stds}); moreover, the majority of our targets do not exhibit variability at the 3-$\sigma$ level.  We are therefore confident that we have not introduced a spurious signal in our data reduction, and our reported photometry is robust.

\subsection{IR Color Variability}

ESHC and ELHC stars are known to exhibit two different types of photometric color 
variability, those which become bluer when fainter (i.e. exhibit a negative color gradient) and those which become redder when fainter (i.e. exhibit a positive color gradient).  The bluer when fainter phenomenon was originally suggested to arise from 
nebular scattered light becoming a more dominant component when the central star itself 
exhibited photometric fading \citep{lam99,dew02}.  However \citet{dew05} noted that the 
bluer when fainter phenomenon could also arise from simple variable continuum bf-ff radiation from classical Be stars, and that this was a simpler explanation than that presented by others.  The redder when fainter stars have been suggested to be young pre-main sequence stars which are variably attenuated by inhomogeneous dust clouds in their 
circumstellar matter \citep{lam99,dew02,dew05}.

Most of the ELHC stars we identify as having 3-$\sigma$ near-IR variability exhibit a grey color variation (ELHC 7,12,21; Table \ref{photdiff}).  ELHC 10 only exhibits 3-$\sigma$ variability in H-band, but if one considers its 2-$\sigma$ J and Ks band variations to be real then it also exhibits a grey color variation.  The only star which exhibits a significant color change is ELHC 3 (Table \ref{photdiff}), which 
exhibited significantly less Ks-band flux compared to other filters in our data when compared with archival observations \citep{dew05}.

\subsection{2 Color Diagram}

\begin{figure}
\begin{center}
\includegraphics[scale=0.7]{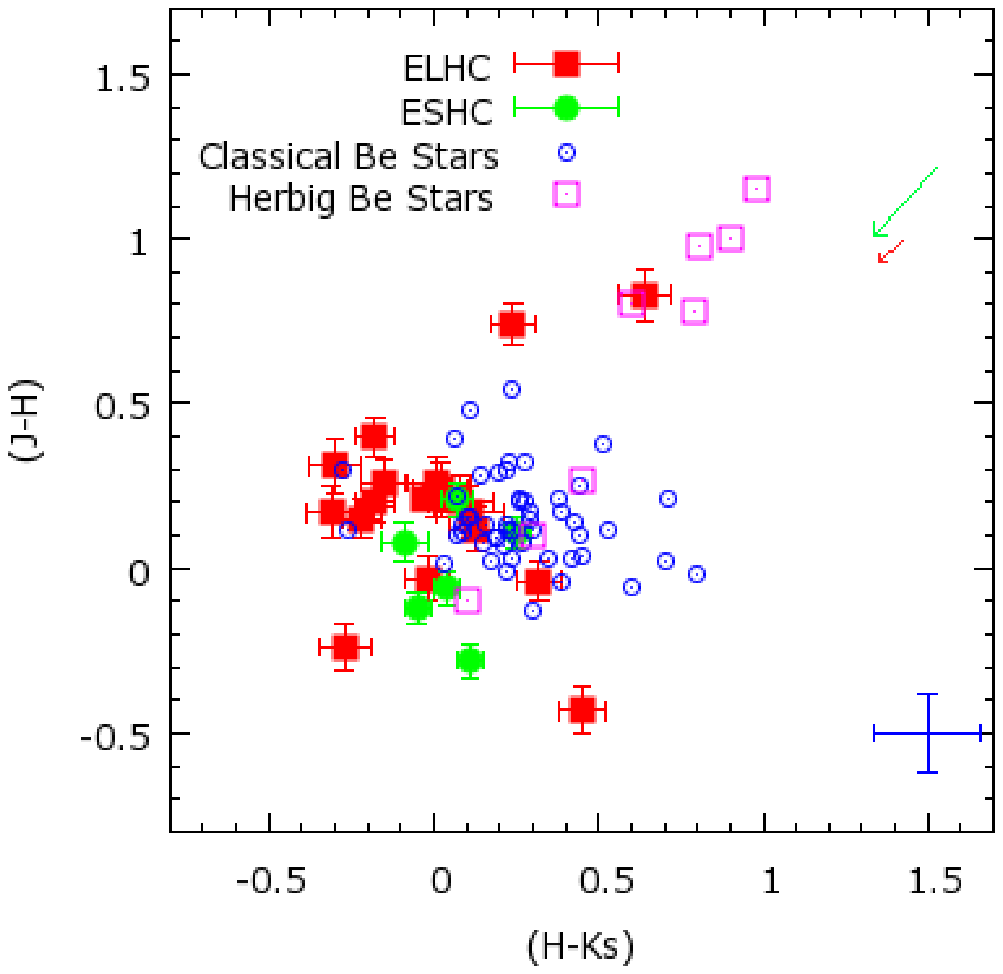}
\caption{Our new IR photometric observations of ESHC and ELHC stars are plotted on an 
IR 2-color diagram (filled circles for ESHC stars, filled squares for ELHC stars).  We also plot the 2MASS IR colors of confirmed classical Be stars in the SMC and LMC by \citet{wis07} as open circles, and the colors of Galactic HAeBe stars by \citet{hil92} as open squares.  The arrows (red for LMC, green for SMC) denote the effects of dust extinction for A$_{V}$ = 2.5$^{m}$. \label{2cdfig}}
\end{center}
\end{figure}

We construct an IR 2-color diagram (2-CD) for 27 of the 28 ESHC and ELHC stars for which we report new IR photometry in this paper, shown in Figure \ref{2cdfig}.  For comparison, we also plot the 2MASS IR colors of sources confirmed to be classical Be stars in the SMC and LMC by \citet{wis07}, i.e. sources which exhibited polarization Balmer jumps or electron scattering signatures.  A selection of Galactic HAeBe stars from \citet{hil92} is also shown.  We find that the near-IR properties of SMC/LMC classical Be stars share many similarities with SMC/LMC candidate HAeBe stars and even some overlap with Galactic HAeBe stars, a phenomenon which has been similarly noted by \citet{dew03} and \citet{dew05}.  This near-IR degeneracy is likely 
influenced by the lower metallicity content of the SMC/LMC, which might reduce the IR excess of HAeBe stars \citep{dew03}, and also lead to potentially stronger free-free IR excesses of classical Be stars in the SMC/LMC as compared to our Galaxy \citep{bon09,bon10}.  Though our sample size is much smaller, we see no distinct evidence in Figure \ref{2cdfig} that the IR excesses of the ESHCs are systematically larger than those of the ELHCs, which might be expected if the sources were classical Be stars \citep{bon10}.

\subsection{Optical Spectroscopy}

\begin{figure}
\begin{center}
\includegraphics[scale=0.35]{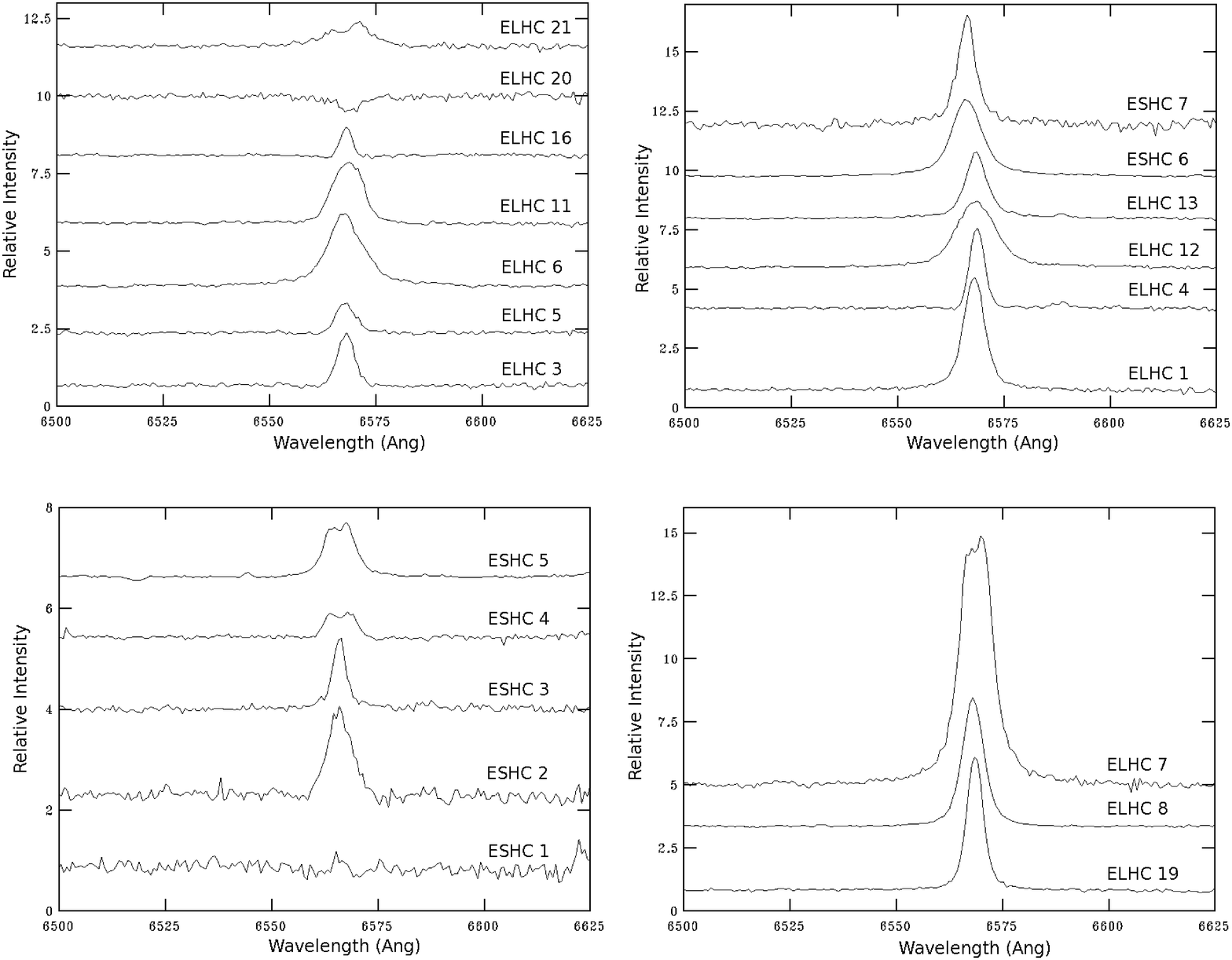}
\caption{The continuum normalized H$\alpha$ line profile for each of the 21 ELHC and ESHC stars we observed with Hydra at the CTIO 4m are presented.  Each spectrum has been shifted vertically to enable details of the line profile to be seen.  Note that we have not doppler corrected the spectra to account for the different average velocities of the SMC versus LMC.  Strong emission is seen in all sources except for ELHC 20, which exhibits very weak emission at the center of its underlying photospheric absorption line, and ESHC 1, whose emission simply fills in its photospheric absorption line.  \label{specfig}}
\end{center}
\end{figure}

We analyzed the basic properties of the H$\alpha$ and H$\beta$ lines in our continuum normalized spectroscopic data, taken nearly contemporaneously with our near-IR photometry, 
to help assess the IR variability and colors we observed.  The full width at half maximum (FWHM) and EW of H$\alpha$, ratio of the peak-to-continuum strength of H$\alpha$, and basic line profile shapes of H$\alpha$ and H$\beta$ are detailed in Table \ref{spec}, while H$\alpha$ line profiles are plotted in Figure \ref{specfig}.  
For reference, nebular features in non sky subtracted imagery exhibited a FWHM of $\sim$3\AA.  When available, archival spectroscopic measurements for some of these sources are also compiled in Table \ref{specfig}.  We comment on individual sources below.

\begin{itemize}

\item \textbf{ELHC 1, 11; ESHC 3, 6, 7} Our new observations of ELHC 1, ELHC 11, ESHC 6, and ELHC 7 all exhibit strong emission in H$\alpha$, with EWs ranging from -6.8 (ESHC 3) to -34.3 (ELHC 1), FWHMs ranging from 4.0 \AA\ (ESHC 3) to 9.1 \AA\ (ESHC 6), and peak-to-continuum ratios ranging from 2.4 (ESHC 3) to 5.7 (ELHC 1).  With the exception of ESHC 3, whose H$\beta$ emission merely fills in the underlying photospheric absorption line, each source also exhibits emission at H$\beta$ which rises above the continuum level.  ELHC 11 was noted in Section \ref{confuse} to suffer from source confusion owing to a star $\sim$1$\farcs$5 away.  Given the 2$\farcs$0 fiber width of the Hydra spectrograph, we likely captured signal from both ELHC 11 and its nearby neighbor.  Its emission line characteristics listed in Table \ref{spec} should be viewed with extreme caution. 

\item \textbf{ESHC 4, 5} Our new observations of ESHC 4 and ESHC 5 show each source exhibits double-peaked H$\alpha$ emission with peak-to-peak separations of $\sim$4.3 \AA\ (ESHC 4) and $\sim$3.3 \AA\ (ESHC 5).  While ESHC 4 exhibits evidence of weak H$\beta$ emission which does not emerge above the continuum level, ESHC 5 exhibits no obvious signatures of emission at H$\beta$.

\item \textbf{ELHC3}  Although earlier epoch (1994 Jan) spectroscopy by \citet{lam99} exhibits an H$\alpha$ EW and FWHM a factor of two stronger than that found in our 2005 Jan 21 data, we observe a stronger ratio of the peak-to-continuum H$\alpha$ emission in 2005 (see e.g. Table \ref{spec}).  The resolution of the 1994 data at H$\alpha$ (R $\sim$2000) is similar to the resolution of our data (R $\sim$2200), indicating these variations are not instrumental artifacts.  These results indicate that the H I line core flux may have increased from 1994 to 2005, while the wing flux decreased.

\item \textbf{ELHC 4}  The ratio of the H$\alpha$ peak-to-continuum emission is much stronger in our 2005 Jan 21 data (4.4) than that we estimated for the 1994 Aug data ($\sim$2.2) of \citet{lam99}.  The resolution of the 1994 data at H$\alpha$ (R $\sim$1300) is somewhat less than our data (R $\sim$2200), which could contribute to some of the lower peak-to-continuum ratio seen in the earlier epoch data.  Moreover, ELHC 4 was noted in Section \ref{confuse} to suffer from source confusion owing to a star $\sim$1$\farcs$2 away.  We likely captured signal from both ELHC 4 and its nearby neighbor given the 2$\farcs$0 fiber width of the Hydra spectrograph; hence, emission line characteristics listed in Table \ref{spec} should be viewed with extreme caution. 

\item \textbf{ELHC 5} Comparison of our 2005 Jan 21 data and 1994 Aug data obtained by \citet{lam99} suggests that measured H I line differences (Table \ref{spec}) predominantly occured the line wings.

\item \textbf{ELHC 6} The H$\alpha$ EW strength, FWHM, and peak-to-continuum ratio between our 2001 Jan 21 data and archival spectral from 1994 Aug \citep{lam99} clearly differ.  Both data were obtained at similar resolutions (R $\sim$2000 vs $\sim$2200), hence resolution likely plays little role in the differences observed.  We caution however that ELHC 6 suffers from clear evidence of source confusion (Section \ref{confuse}); hence, emission line characteristics listed in Table \ref{spec} should be viewed with extreme caution. 

\item \textbf{ELHC 7} We observe a double-peaked H$\alpha$ emission line profile which has a similar EW and FWHM to that reported in 1994 Aug spectra (\citealt{lam99}).  \citet{lam99} do not report the H$\alpha$ profile to be double-peaked in their data; however, their lower resolution (R $\sim$1300) might contribute to this non-detection especially given the peak-to-peak separation we see in our data ($\sim$2.3 \AA ).

\item \textbf{ELHC 8} \citet{dew05} reported a similar H$\alpha$ EW in their low resolution (R $\sim$400) spectrum obtained on 2000 January 8 as that measured in our 2005 Jan 21 data.  The H$\alpha$ FWHM in the 2000 data (13.5 \AA ) appears to be below the resolving power of the spectrograph used by \citet{dew05}; hence, comparisons of the change in line width are not possible.

\item \textbf{ELHC 12} \citet{dew05} obtained a low resolution (R $\sim$450) spectrum of ELHC 12 on 1998 January 6, shortly before the near-IR photometry they obtained on 1998 January 19-20.  Altough they didn't report H$\alpha$ EW or FWHM measurements for these data, we did estimate the ratio of the peak-to-continuum emission of their H$\alpha$ profile to be $\sim$3.1, which is similar to the value of 3.8 we derived from our 2005 January 21 data, especially given the difference in the resolution of the two different data sets.  Based on these available data, we find no clear evidence that the H$\alpha$ spectroscopic properties of ELHC 12 were substantially different between the 1998 January epoch observation and the 2005 January epoch observation.

\item \textbf{ELHC 13} Our 2005 Jan 21 spectrum exhibited a smaller H$\alpha$ EW than that observed in a low resolution (R $\sim$ 400) spectra obtained on 2000 March 9 by \citet{dew05}.  The H$\alpha$ FWHM in the 2000 data (14.4 \AA ) appears to be below the resolving power of the spectrograph \citet{dew05} used; hence, comparisons of the change in line width are not feasible and the observed differences in the H$\alpha$ peak-to-continuum ratio might be influenced by the large difference in resolution of the two datasets.

\item \textbf{ELHC 16} \citet{dew05} obtained a low resolution (R $\sim$450) spectrum on 1998 January 6 and reported weak H$\alpha$ emission, which filled in the underlying photospheric absorption line.  By contrast, we observe emission in H$\alpha$ which peaks 1.9x above the continuum, as well as weak emission in H$\beta$ which partially fills in the absorption line in our 2005 January 21 data.  The H$\alpha$ FWHM of our observation (3 \AA ) is not broader than our instrumental resolution; however, we see no strong evidence of residual nebular emission in our spectra following sky subtraction, so we suggest we are seeing emission which is circumstellar in origin.

\item \textbf{ELHC 19} \citet{dew05} observed this star at low resolution (R $\sim$450) on 1998 January 6.  While they didn't tabulate H$\alpha$ line strengths or widths, we did estimate the ratio of the peak-to-continuum emission in their H$\alpha$ profile to be $\sim$2, which was substantially smaller than the ratio of 6.3 we observed in our 2005 January 21 data.  Although the different resolution of the data likely contributes to some of the observed differences, these results suggest ELHC 19 exhibited significantly stronger emission in 2005 as compared to 1998.

\item \textbf{ELHC 20} We only detected weak evidence of emission in the core of the H$\alpha$ photospheric absorption line of ELHC 20 in our 2005 January 21 data, which qualitatively matches the filled-in absorption line observed in R $\sim$450 spectra by \citet{dew05} on 1998 January 6.

\item \textbf{ELHC 21} \citet{dew05} obtained a R $\sim$1000 spectrum of ELHC 21 on 1998 January 8 (nearby the epoch of their near-IR photometry of the source) and found H$\alpha$ to have a slightly larger FWHM than observed in our 2005 January 21 spectrum, although the line's peak-to-continuum line ratio were similar in both epochs.  The peak-to-peak separation in the double-peaked emission line profile we observe, $\sim$6.7 \AA, is larger than the resolution of the 1998 epoch data; hence, it is difficult to assess whether the morphology of this line has evolved as a function of time.  Based on these available data, we find no clear evidence that the H$\alpha$ spectroscopic properties of ELHC 21 were substantially differenent between the 1998 January epoch observation and the 2005 January epoch observation.

\item \textbf{ESHC 1} The H$\alpha$ EW, FWHM, and peak-to-continuum strength were all larger in the spectrum of ESHC 1 obtained by \citet{bea01} on 1998 August 18 than in our 2005 January 21 data.  The overall emission strength of ESHC 1 therefore seems to have decreased from 1998 to 2005.

\item \textbf{ESHC 2} We observed ESHC 2 to have a weaker H$\alpha$ EW And FWHM than seen in archival observations made on 1998 August 18 by \citet{bea01}.  The peak-to-continuum line strength ratio in each data were similar however, suggesting that the line profile morphology of ESHC 2 has slightly changed between these two epochs, with emission from the line wings decreasing from 1998 to 2005.
\end{itemize}

\section{Discussion}

Classical Be stars can share many similar observational features with HAeBe star and are likely to be more prevalent in the Galaxy and SMC/LMC, given the short pre-main sequence lifetimes of intermediate-mass stars.  Thus, establishing that ESHC and ELHC stars are in fact pre-main sequence objects requires clear evidence that their observed behavior is inconsistent with that expected from classical Be stars.  Figure \ref{2cdfig} illustrates that the IR colors of known SMC/LMC classical Be stars share common parameter space both with many ESHC/ELHC stars and with some Galactic HAeBe stars.   Moreover, \citet{dew05} show that this degeneracy persists after corrections are made to deredden the colors of Galactic HAeBe stars to attempt to match the lower metallicity content of the Magellanic Clouds.  Thus the magnitude of the near-IR excess associated with ESHC and ELHC stars alone is insufficient to conclusively determine whether these objects are HAeBe stars or classical Be stars.       

Some aspects of the magnitude and time-scale of the photometric variability exhibited by HAeBe stars and classical Be stars can be used to help identify the nature of ELHC and ESHC stars.  Classical Be stars are known to aperiodically exhibit dramatic events whereby they completely lose their disks or regenerate a new disk from a disk-less stage (normal B to Be to normal B transitions; \citealt{und82,mcs09,wis10,dra11}).  During these events the stars optical photometry has been observed to change $\sim$0.5 \citep{bjo02} to $\sim$0.7 magnitudes \citep{hum98,gan02,mir03}.  J,H,K-band brightness variations of $<$0.5 - $\sim$0.95 magnitudes have been reported by \citet{ash84} and \citet{dou94} over time-scales of hundreds of days to tens of years; the largest of these IR variations likely correspond to large disk loss/disk-renewal events.  \citet{dou94} suggest that a $\sim$0.5 magnitude variation corresponds to a change in the density of an optically thin disk of a factor of $\sim$2-3.  Large-magnitude, long-term changes in the strength of H$\alpha$ emission and intrinsic polarization are also observed during disk-loss/disk-renewal events \citep{wis10,dra11}.  Some HAeBe stars are also known to exhibit large ($\sim$1 magnitude) IR photometric variability, which sometimes correlates with optical photometric variations and other times is uncorrelated, on time-scales as short as 1-2 days \citep{eir02}.  The origin of the photometric variation in these UX Ori objects has been suggested to be variable obscuration by dust clumps \citep{gri91, wat98}, although alternate mechanisms such as variable accretion have also been suggested \citep{her99}.  UX Ori events also produce significant enhancements in the observed linear polarization, albeit over short time-scales \citep{gri91}.

Given our new high precision near-IR photometry of 27 of 28 ESHCs and ELHCs and new moderate resolution H$\alpha$ and H$\beta$ spectroscopic observations of 21 of 28 of these sources, we now discuss how our new data add additional constraints on the evolutionary status of ESHC and ELHC stars.

\subsection{Source Confusion}

ELHC 6 (this work) and ELHC 18 (this work and \citealt{dew05}) have now each been confirmed to be comprised of two sources; hence, we suggest that previous efforts
to identify and characterize the optical/IR variability and IR excess of these sources should be treated with some caution.  We found ELHC 4 and ELHC 11 had secondary sources located 1$\farcs$2-1$\farcs$5 from each star; hence, previous photometric observations which had poorer spatial resolution (or seeing) than these values may suffer from contamination.  We find no support for possible source confusion for either ELHC 5 or ELHC 8, which contrasts with the suggestions of previous works \citep{dew05,wis06}, and therefore we suggest they continue to be classified as candidate HAeBe targets.

\subsection{The Nature of ELHC 7, ELHC 12, and ELHC 21} \label{nature}

We detected evidence of photometric variability at the 3-$\sigma$ level in one IR filter for ELHC 3 (Ks-band) and ELHC 5 (J-band, though H and Ks filters also exhibit variability at the 2-$\sigma$ level).  ELHC 21, ELHC 7, and ELHC 12 exhibited variability in all 3 filters.  The large ($\sim$1 magnitude) IR photometric brightening observed in both ELHC 7 and ELHC 12 is at the extreme end of IR photometric variability observed in classical Be stars.   ELHC 7 exhibited a very large H$\alpha$ emission line in both 1994 and 2005; hence, it is plausible that the system also exhibited significant H$\alpha$ emission at the earlier epoch of IR photometric observations (January 1998).  Moreover, ELHC 7 has previously been classified as a UX Ori variable \citep{dew05}.  Hence, the IR variability we observe is completely consistent with the classification of this star as a HAeBe star.  ELHC 12 also exhibited a H$\alpha$ emission line with a very similar peak-to-continuum ratio near the epoch of its 1998 IR photometric observation ($\sim$3.1) as near the epoch of its 2005 IR photometric observation (3.8).  Although we do not have information about the strength of the EW and FWHM of H$\alpha$ for the archival 1998 data, we suggest that there is no clear evidence to indicate that the H$\alpha$ emission line was substantially different between the two epochs of observations.  We are not aware of any classical Be star model which would predict the magnitude of near-IR flux increase we see for ELHC 12, and corresponding lack of strong enhancement in the H$\alpha$ profile, owing to a disk building event.  We therefore suggest that the IR photometric variability we observe for ELHC 12 might indicate that the system is a UX Ori-type HAeBe star.

The IR brightening we observe for ELHC 21, $\sim$0.5 magnitudes, is within the range of typical variability observed at near-IR wavelengths for both HAeBe stars and classical Be stars.  If ELHC 21 is a classical Be star, this level of variability could indicate a factor of $\sim$2-3 change in the density of the disk \citep{dou94}.  Comparison of the ratio of the peak-to-continuum line strength of the H$\alpha$ spectra obtained near the time of the 1998 epoch and 2005 epoch IR photometry reveals little evidence of substantial changes.  The region of a Be star's disk responsible for producing H$\alpha$ emission is known to be predominantly located at larger radii than that which produces near-IR emission \citep{wi07b,car11}.  Thus the $\sim$0.5 magnitude IR brightening we observe in our 2005 data could indicate that a significant, recent (inner) disk building event occurred, which had not yet manifested itself in diagnostics more sensitive to the outer disk region (H$\alpha$ spectroscopy).  Alternatively, we can not rule out that the moderate-scale variability we have observed could also arise from circumstellar dust clouds which attenuate light from the central source (see e.g. \citealt{dew05}), and therefore indicate the system is a HAeBe star.  Higher cadence, contemporaneous optical and IR observations of ELHC 21 should be pursued to differentiate between these two scenarios.

\subsection{IR Color Variability}

In contrast to the observed optical variability of ELHC stars, which exhibits either bluer when fainter (negative color gradient) or redder when fainter (positive color gradient) colors, most of the IR variability we observe is grey.  The color effects of nebular scattering, dust reddening, and variable bf-ff emission should all be small at near-IR wavelengths, so it is perhaps not unexpected that the variability we observe 
is grey.  We note that the sources which exhibit the greatest IR variability (ELHC 7, 12, and 21) were found to either have positive optical color gradients (redder when fainter) or grey optical color gradients \citep{dew05}.  Variable dust cloud obscuration has been suggested as the most likely explanation for this optical behavior \citep{dew05}, which is consistent with our previous interpretation for the origin of the IR variability we observe in these sources.  Two ELHC stars which exhibit bluer when fainter optical colors were also identified to be variable in one IR filter in our 
data (ELHC 3 and ELHC 5); however, the single bandpass IR variability detection precludes us from drawing strong conclusions on the color variability of these systems.

\subsection{Optical Spectroscopic Variability}

The new H$\alpha$ spectroscopy presented in this paper enables us to investigate the spectroscopic variability of numerous ESHC and ELHC stars which have earlier epoch spectroscopic data in the literature.  By analyzing the differences in the EW, FWHM, and peak-to-continuum ratio, we've been able to identify evidence of strengthening line core and/or wing emission in ELHC 3, ELHC 16, and ELHC 19 and decreasing line core and/or wing emission in ELHC 5, ELHC 13, ESHC 1, and ESHC 2.  We detect modest or minimal changes in the H$\alpha$ emission strength in ELHC 7, ELHC 8, ELHC 20, and ELHC 21.  As the population of ESHC and ELHC stars were initially identified on the basis of their observed optical photometric variability, it is not unexpected that we have detected a variety of changes in the appearance of each star's H$\alpha$ line emission.  The emission strength and line profiles of both HAeBe stars \citep{pog94,rei96,vie99} and classical Be stars \citep{por03,ste09,wis10} are known to exhibit variability owing to changes in their circumstellar environments; hence, the presence of this variability alone is not a good discriminant between the two types of stars.

\subsection{New IR Photometry and Optical Spectroscopy}

We have reported near-IR photometry for ESHC 2 and ESHC 3 (Table \ref{phot}).  Both the (J-H) and (H-Ks) colors are close to zero or marginally blue.  \citet{dew03} adopted an optically thin dust scaling algorithm to estimate how the colors of Galactic HAe and HBe stars would be altered if these sources were in the dust poor SMC, and in Figure 17 of their paper they show a clear degeneracy between the IR colors of classical Be stars and HAe and HBe stars.  This degeneracy incorporates the range of (J-H) and (H-Ks) colors we report for ESHC 2 and ESHC 3 (see Figure \ref{2cdfig}); hence, the colors of our new observations do not permit us to differentiate the nature of these objects.  Moreover, as no pre-existing IR photometry is available, we can not measure the variability (or lack thereof) of these sources for use as a discriminant.

We have also reported H$\alpha$ spectroscopy for ELHC 1 and ESHC 3, 4, 5, 6, and 7 and tabulated H$\alpha$ line strengths and properties for ELHC 11, 12, 16, 19, 20, and 21.  All sources exhibited H$\alpha$ in emission, although line strengths varied greatly from sources which exhibited small hints of emission in the core of their underlying photospheric absorption lines (ELHC 20) to strong emission with EWs between -33 - -34 (ELHC 1, ESHC 6).  The moderate resolution (R $\sim$2200) of our data enabled us to identify evidence of double-peaked line profiles in ELHC 21, ESHC 4, and ESHC 5.  These new data do not provide definitive differentiation of the classification of ESHC and ELHC stars as HAeBe versus classical Be stars.  However, as seen in Section \ref{nature}, having quality near-contemporaneous optical spectroscopic and IR photometric observations in the literature can provide invaluable insight into the time evolution and hence origin of these objects.

\section{Summary}
We obtained IR photometric observations of 27 of the 28 ESHC and ELHC stars identified to date, and near-contemporaneous optical spectroscopic observations of 21 of these 28 sources and found: 

\begin{itemize}
\item Earlier epoch observations of ELHC 4, 6, 11, and 18, obtained with facilities with lower angular resolution than our study, likely suffer from source confusion.  For these stars, the photometric variability and colors, hence classification as candidate HAeBe stars, are therefore suspect.
\item ELHC 5, 7, 12, and 21 exhibit statistically significant photometric variability in 1 or more near-IR filters.  These photometric variations exhibit a 
grey color difference as compared to earlier epoch observations.
\item ELHC 7 and ELHC 12 exhibit $\sim$1 magnitude IR brightness enhancements in our data, but their H$\alpha$ emission-line strengths do not seem to have 
appreciably changed.  We suggest these observational phenomenon could be produced by dust clouds which variably attenuate the observed brightness of these systems.  ELHC 7 has previously been suggested to be a UX Ori-type variable, which our observations support.  Our suggestion that ELHC 12 might similarly be a UX Ori-type variable has not been previously reported in the literature.  
\item ELHC 21 exhibits a $\sim$0.5 magnitude IR brightness enhancement in our data, with little corresponding change in its H$\alpha$ emission strength.  If ELHC 21 is a Be star, these observations could indicate that the system experienced a major injection of new material into its inner disk region.  We note however that we can not strictly rule out that the observed behavior was caused by variable obscuration by dust clouds surrounding a HAeBe star.
\item We present the first near-IR photometry for ESHC 2 and 3 and first H$\alpha$ spectroscopy for ELHC 1 and ESHC 3, 4, 5, 6, and 7.  Although we detect H$\alpha$ to be in emission in each of these sources, the available data do not allow us to place strict constraints on whether the sources are HAeBe stars or classical Be stars.  
\end{itemize}

\acknowledgments

We thank the referee for providing feedback which helped improve the clarity of this paper.  We also thank the NOAO TAC 
for allocating observing time for this project and JPW thanks NOAO for supporting his travel to 
CTIO.  Portions of this work have been supported in part by NASA NPP NNH06CC03B (JPW), NASA 
GRSP NGT5-50469 (JPW), and NSF AST 08-02230 (JPW).  This research has made use of the SIMBAD database operated 
at CDS, Strasbourg, France, and the NASA ADS system.  

{\it Facilities:} \facility{Blanco}

\newpage
\clearpage
\begin{table}
\begin{center}
\scriptsize
\caption{Coordinates of candidates \label{targetcoords}}
\begin{tabular}{lccc}
\tableline
Star & RA & Dec  & EROS name \\ 
\tableline
ELHC 1 & 	05 18 14.7 & -69 30 16.3	& 01-1603 \\
ELHC 2 &	05 18 32.9 & -69 39 43.8	& 02-111 \\
ELHC 3	 & 05 16 29.6	& -69 17 17.4	& 08-830 \\
ELHC 4	& 05 17 17.6	& -69 21 35.8	& 08-1335 \\
ELHC 5	& 05 18 20.7	& -69 32 40.7	& 01-1414 \\
ELHC 6	& 05 18 18.7	& -69 35 30.0	& 01-1457 \\
ELHC 7	& 05 16 39.2	& -69 20 47.5	& 08-1005 \\
ELHC 8	& 05 17 17.1	& -69 33 35.0	& 01-837 \\ 
ELHC 9	& 05 17 07.0	& -69 33 22.8	& 01-749 \\
ELHC 10	& 05 19 47.8	& -69 39 11.6	& 02-765 \\ 
ELHC 11	& 05 18 35.4	& -69 36 02.8	& 02-915 \\
ELHC 12	& 05 18 50.4	& -69 35 24.8	& 02-999 \\ 
ELHC 13	& 05 18 54.7	& -69 36 35.3	& 02-1047 \\
ELHC 14	& 05 19 54.3	& -69 42 07.4	& 03-14 \\
ELHC 15	& 05 27 22.9	& -69 52 15.3	& 07-1108 \\
ELHC 16	& 05 20 27.5	& -69 35 39.5	& 03-689 \\
ELHC 17	& 05 27 24.8	& -69 51 50.2	& 07-1120 \\
ELHC 18	& 05 27 22.7	& -69 52 02.0	& 07-1098 \\
ELHC 19	& 05 17 11.0	& -69 25 52.3	& 08-649 \\
ELHC 20	& 05 16 22.5	& -69 20 18.1	& 08-842 \\
ELHC 21	& 05 19 44.2	& -69 26 13.9	& 10-430 \\
ESHC 1	& 00 53 02.8	& -73 17 58.8	& sm00103k-4755 \\
ESHC 2	& 00 52 32.8	& -73 17 07.3	& sm00103k-2210 \\
ESHC 3	& 00 52 21.6	& -73 13 32.4	& sm00101l-15299 \\
ESHC 4	& 00 53 56.8	& -73 10 30.0	& sm00101n-14763 \\
ESHC 5	& 00 52 06.4	& -73 06 29.1	& sm00101l-3359 \\
ESHC 6	& 00 54 37.5	& -73 04 56.4	& sm00101n-2402 \\
ESHC 7	& 00 52 52.6	& -73 18 33.3	& sm00103k-6329 \\
\tableline
\tablecomments{The name of each of the ESHC/ELHC targets we observed, along with updated coordinates accurate to 0.1 seconds in 
RA and 0$\farcs$4 in Dec are compiled.}
\end{tabular}
\end{center}
\end{table}

\newpage
\clearpage
\begin{table}
\begin{center}
\scriptsize
\caption{Summary of Observing Fields \label{zps}}
\begin{tabular}{lcccc}
\tableline\tableline
Field Name & Candidates in Field & Zero Point$_{J}$ & Zero Point$_{H}$ & Zero Point$_{Ks}$ \\
\tableline
LMC 1	& ELHC 15, 17, 18 & 21.52	$\pm$ 0.04 & 	22.23 $\pm$ 0.03	& 21.85 $\pm$ 0.04 \\
LMC 2	& ELHC 1, 5, 6, 11, 12, 13	& 21.87	$\pm$ 0.03	& 22.35	$\pm$ 0.03	& 21.79	$\pm$ 0.03 \\
LMC 3	& ELHC 8, 9, 19	& 21.75	$\pm$ 0.03	& 22.15	$\pm$ 0.03	& 21.44	$\pm$ 0.04 \\
LMC 4	& ELHC 3, 4, 7, 20	& 21.85	$\pm$ 0.04	& 22.14	$\pm$ 0.04	& 21.69	$\pm$ 0.04 \\
LMC 5	& ELHC 10, 14, 16	& 21.87	$\pm$ 0.04	& 22.30 $\pm$ 0.04	& 21.82 $\pm$ 0.06 \\
LMC 6	& ELHC 21	& 21.63 $\pm$ 0.03 & 22.04 $\pm$ 0.04 & 21.65	$\pm$ 0.04 \\
LMC 7	& ELHC 2	& 21.64	$\pm$ 0.04	& 21.75 $\pm$ 0.04 & 21.47	$\pm$ 0.04 \\
SMC 1	& ESHC 1, 2, 3, 7	& 21.87	$\pm$ 0.03 & 	22.40	$\pm$ 0.02	& 21.95	$\pm$ 0.02 \\
SMC 2	& ESHC 5	& 21.96	$\pm$ 0.03	& 22.40	$\pm$ 0.02	& 21.94	$\pm$ 0.03 \\
SMC 3	& ESHC 6	& 21.96	$\pm$ 0.03	& 22.42	$\pm$ 0.03	& 21.94	$\pm$ 0.03 \\
HST9106	& \nodata &	21.79	$\pm$ 0.02	& 22.27	$\pm$ 0.02	& 21.85	$\pm$ 0.01 \\
HST9115	& \nodata &	21.90 $\pm$	0.01	& 22.26	$\pm$ 0.08	& 21.99	$\pm$ 0.03 \\
HST9119	& \nodata & 21.84 $\pm$ 0.02	 & 22.31	$\pm$ 0.04	& 21.86	$\pm$ 0.03 \\
HST9121	& \nodata &	21.73	$\pm$ 0.05	& 22.04	$\pm$ 0.05	& 21.66	$\pm$ 0.07 \\
\tableline
\tablecomments{Basic parameters for each of our fields of view, such as the names of the ESHC/ELHC candidates contained within and derived 
photometric zeropoints, are provided. }
\end{tabular}
\end{center}
\end{table}

\newpage
\clearpage
\begin{table}
\begin{center}
\scriptsize
\caption{Summary of Standard Star Observations \label{stds}}
\begin{tabular}{lcccc}
\tableline\tableline
Name & Filter & m$_{this study}$ & m$_{2MASS}$ & $\delta$m \\
\tableline
HST9106	& J & 12.204 $\pm$ 0.018 & 12.129 $\pm$ 0.011 & 0.075 $\pm$ 0.029 \\
\nodata & H & 11.870 $\pm$ 0.020 & 11.852 $\pm$ 0.036 & 0.022 $\pm$ 0.056 \\
\nodata & Ks & 11.774 $\pm$ 0.010 & 11.758 $\pm$ 0.025 & 0.016 $\pm$ 0.035 \\
HST9115	& J & 12.028 $\pm$ 0.010 & 12.017 $\pm$  0.007	& 0.011 $\pm$ 0.017	\\
\nodata & H & 11.755 $\pm$ 0.084 & 11.856 $\pm$ 0.038 & -0.101 $\pm$ 0.122 \\ 
\nodata & Ks & 11.834 $\pm$ 0.030 & 11.803 $\pm$ 0.023	& 0.031 $\pm$  0.053 \\
HST9119	& J & 12.089 $\pm$ 0.018 & 12.066 $\pm$  0.018	& 0.023 $\pm$ 0.036	\\
\nodata & H & 11.798 $\pm$ 0.037	& 11.794 $\pm$  0.039	& 0.004 $\pm$ 0.076	\\
\nodata & Ks & 11.732 $\pm$ 0.029	& 11.775 $\pm$ 0.021	& -0.043 $\pm$ 0.050 \\
HST9121	& J & 11.680 $\pm$ 0.050	& 11.685 $\pm$ 0.026	& 0.005 $\pm$ 0.076	\\
\nodata & H & 11.373 $\pm$ 0.056	& 11.413 $\pm$ 0.019 & -0.040 $\pm$ 0.075	\\ 
\nodata & Ks & 11.363 $\pm$ 0.071	& 11.366 $\pm$ 0.012	& -0.003 $\pm$ 0.083 \\
\tableline
\tablecomments{The absolute photometry values that we derived for our standard stars are listed, along with the difference between these values and 
archival 2MASS values.}
\end{tabular}
\end{center}
\end{table}

\newpage
\clearpage
\begin{table}
\begin{center}
\scriptsize
\caption{Photometry and Colors of ESHCs and ELHCs \label{phot}}
\begin{tabular}{lccccc}
\tableline\tableline
Name & J & H & Ks & (J-H) & (H-Ks) \\
\tableline
ELHC 1	& 15.25 $\pm$ 0.03 & 15.05 $\pm$ 0.03 & 15.23 $\pm$ 0.03	& 0.20 $\pm$ 0.06 & -0.18 $\pm$ 0.06 \\
ELHC 2	 & 15.50 $\pm$ 0.04	& 15.74 $\pm$ 0.04	& 16.01 $\pm$ 0.04	& -0.24 $\pm$ 0.07	& -0.27 $\pm$ 0.08 \\
ELHC 3	 & 16.45 $\pm$ 0.04	& 16.14 $\pm$ 0.04	& 16.44 $\pm$ 0.04	& 0.31 $\pm$ 0.08	& -0.3 $\pm$ 0.08 \\
ELHC 4	 & 15.06 $\pm$ 0.04	& 14.82 $\pm$ 0.04	& 14.80 $\pm$ 0.04	& 0.24 $\pm$ 0.08 &	0.02 $\pm$ 0.08 \\
ELHC 5	 & 16.13 $\pm$ 0.03	& 15.98 $\pm$ 0.03	& 16.20 $\pm$ 0.03	& 0.15 $\pm$ 0.06	& -0.22 $\pm$ 0.06 \\
ELHC 6	  & \nodata & \nodata & \nodata & \nodata & \nodata \\	 	 	 	 
ELHC 7	& 16.79 $\pm$ 0.04	& 15.96 $\pm$ 0.04	& 15.32 $\pm$ 0.04	& 0.83 $\pm$ 0.08	& 0.64 $\pm$ 0.08 \\
ELHC 8	 & 15.35 $\pm$ 0.03	& 15.39 $\pm$ 0.03	& 15.07 $\pm$ 0.04	& -0.04 $\pm$ 0.06	& 0.32 $\pm$ 0.07 \\
ELHC 9	 & 16.10 $\pm$ 0.03	& 15.36 $\pm$ 0.03	& 15.12 $\pm$ 0.04	& 0.74 $\pm$ 0.06	& 0.24 $\pm$ 0.07 \\
ELHC 10	& 13.19 $\pm$ 0.04	& 12.93 $\pm$ 0.04	& 12.92 $\pm$ 0.06	& 0.26 $\pm$ 0.08	& 0.01 $\pm$ 0.10 \\
ELHC 11	& 15.08 $\pm$ 0.03	& 14.68 $\pm$ 0.03	& 14.85 $\pm$ 0.03	& 0.40 $\pm$ 0.06	& -0.18 $\pm$ 0.06 \\
ELHC 12	& 15.46 $\pm$ 0.03	& 15.24 $\pm$ 0.03 &	15.25 $\pm$ 0.03	& 0.22 $\pm$ 0.06	& -0.01 $\pm$ 0.06 \\
ELHC 13	& 15.34 $\pm$ 0.03	& 15.13 $\pm$ 0.03	& 15.16 $\pm$ 0.03	& 0.21 $\pm$ 0.03	& -0.03 $\pm$ 0.03 \\
ELHC 14	& 15.68 $\pm$ 0.04	& 15.48 $\pm$ 0.04	& 15.40 $\pm$ 0.06	& 0.20 $\pm$ 0.08	&0.08 $\pm$ 0.10 \\
ELHC 15	& 14.62 $\pm$ 0.04	& 15.05 $\pm$ 0.03	& 14.60 $\pm$ 0.04	& -0.43 $\pm$ 0.07	&0.45 $\pm$ 0.07 \\
ELHC 16	& 15.93 $\pm$ 0.04	& 15.76 $\pm$ 0.04	& 15.65 $\pm$ 0.06	& 0.17 $\pm$ 0.08	&0.11 $\pm$ 0.10 \\
ELHC 17	& 14.11 $\pm$ 0.04	& 14.14 $\pm$ 0.03	& 14.16 $\pm$ 0.04 &	-0.03 $\pm$ 0.07	&-0.02 $\pm$ 0.07 \\
ELHC 18	& 15.40 $\pm$ 0.04	& 15.14 $\pm$ 0.03	& 15.29 $\pm$ 0.04 & 0.26 $\pm$ 0.07	&-0.15 $\pm$ 0.07 \\
ELHC 19	& 16.29 $\pm$ 0.03	& 16.31 $\pm$ 0.03	& \nodata & 	-0.02 $\pm$ 0.06	& \nodata \\
ELHC 20	& 16.11 $\pm$ 0.04	& 15.94 $\pm$ 0.04	& 16.25 $\pm$ 0.04	& 0.17 $\pm$ 0.08	&-0.31 $\pm$ 0.08 \\
ELHC 21	& 15.10 $\pm$ 0.03	& 14.98 $\pm$ 0.04	& 14.85 $\pm$ 0.04	& 0.12 $\pm$ 0.07	&0.13 $\pm$ 0.08 \\
ESHC 1	& 15.28 $\pm$ 0.03	& 15.40 $\pm$ 0.02	& 15.44 $\pm$ 0.02	& -0.12 $\pm$ 0.05	&-0.05 $\pm$ 0.04 \\
ESHC 2	& 17.29 $\pm$ 0.03	& 17.57 $\pm$ 0.02	& 17.46 $\pm$ 0.02	& -0.28 $\pm$ 0.05	&0.11 $\pm$ 0.04 \\ 
ESHC 3	& 16.48 $\pm$ 0.03	& 16.54 $\pm$ 0.02	& 16.50 $\pm$ 0.02	&-0.06 $\pm$ 0.05	 &0.04 $\pm$ 0.04 \\
ESHC 5	& 13.66 $\pm$ 0.03	& 13.45 $\pm$ 0.02	& 13.38 $\pm$ 0.03	&0.21 $\pm$ 0.05	&0.07 $\pm$ 0.05 \\
ESHC 6	& 14.86 $\pm$ 0.03	& 14.78 $\pm$ 0.03	& 14.87 $\pm$ 0.03	&0.08 $\pm$ 0.06	&-0.09 $\pm$ 0.07 \\
ESHC 7	& 14.29 $\pm$ 0.03	& 14.18 $\pm$ 0.02	& 13.92 $\pm$ 0.02	&0.11 $\pm$ 0.05	&0.25 $\pm$ 0.04 \\
\tableline
\tablecomments{The absolute photometry values for all of the known ESHCs and ELHCs is given, along with their (J-H) and (H-Ks) colors.  We detected 2 sources at the coordinates of ELHC 6 given by \citet{dew02}, and discuss the photometry of this composite source in Section \ref{confuse}.}
\end{tabular}
\end{center}
\end{table}

\newpage
\clearpage
\begin{table}
\begin{center}
\scriptsize
\caption{Photometry Variability of ELHCs \label{photdiff}}
\begin{tabular}{lccc}
\tableline\tableline
Name & $\Delta$J & $\Delta$H & $\Delta$Ks  \\
\tableline
ELHC 1	& 0.14 $\pm$ 0.13	& 0.08 $\pm$ 0.18	& 0.35 $\pm$ 0.18 \\
ELHC 2	 & -0.36 $\pm$ 0.14 &	-0.09 $\pm$ 0.19	& \nodata \\
ELHC 3	 & 0.26 $\pm$ 0.14	& 0.11 $\pm$ 0.14	& 0.64 $\pm$ 0.19 \\
ELHC 4	& -0.29 $\pm$ 0.54	& -0.14 $\pm$ 0.09 & -0.27 $\pm$ 0.14 \\
ELHC 5	 & -0.44 $\pm$ 0.13	& -0.48 $\pm$ 0.18	& -0.29 $\pm$ 0.18 \\	
ELHC 7	 & -0.81 $\pm$ 0.19	& -1.03 $\pm$ 0.19	& -0.64 $\pm$ 0.09 \\
ELHC 8 &	-0.19 $\pm$ 0.08	& -0.07 $\pm$ 0.08	& -0.33 $\pm$ 0.09 \\		
ELHC 10	& -0.23 $\pm$ 0.09	& -0.34 $\pm$ 0.09	& -0.22 $\pm$ 0.11 \\		
ELHC 12	& -0.97 $\pm$ 0.13	& -1.12 $\pm$ 0.18	& -1.10 $\pm$ 0.23 \\
ELHC 14	& 0.13 $\pm$ 0.09	& -0.03 $\pm$ 0.14	& -0.16 $\pm$ 0.16 \\	
ELHC 17	& -0.16 $\pm$ 0.09	& -0.03 $\pm$ 0.13	& 0.13 $\pm$ 0.14 \\
ELHC 18	& -0.80 $\pm$ 0.19	& -1.04 $\pm$ 0.18	& -0.99 $\pm$ 0.24 \\
ELHC 19	& -0.06 $\pm$ 0.18	& 0.08 $\pm$ 0.23 & \nodata \\	
ELHC 20	& 0.13 $\pm$ 0.19	& 0.03 $\pm$ 0.24 & \nodata \\	
ELHC 21	& -0.52 $\pm$ 0.13	& -0.51 $\pm$ 0.14	& -0.54 $\pm$ 0.14 \\

\tableline
\tablecomments{The difference between our near-IR absolute photometry and that compiled by \citet{dew05}, with negative numbers indicating the source was brighter at the epoch of our observations versus the literature.}
\end{tabular}
\end{center}
\end{table}

\newpage
\clearpage
\begin{table}
\begin{center}
\scriptsize
\caption{Spectroscopy of ESHCs and ELHCs \label{spec}}
\begin{tabular}{lcccccc}
\tableline\tableline
Name & Date & FWHM$_{H\alpha}$ & EW$_{H\alpha}$ & H$\alpha$ Profile & H$\beta$ Profile  & SNR \\
\nodata & \nodata & \AA\ & \AA\ & strength, profile & profile & \nodata \\
\tableline
ELHC 1	& 2005 Jan 21 & 6.4 &  -34.3 & 5.7, eac$^{1}$ & eac  & 25 \\
ELHC 3 & 1994 Aug$^{6}$ & 9.2 & -15 & $\sim$2$^{5}$, eac & filled abs line & 50 \\
ELHC 3 & 2005 Jan 21 & 4.7 & -8.3 & 2.75, eac & weak eac  & 26 \\
ELHC 4 & 1994 Aug$^{6}$ & 5.3 & -16 & $\sim$2.2$^{5}$, filled abs line & filled abs line & 50 \\
ELHC 4 & 2005 Jan 21 & 4.0$^{4}$ & -14.9$^{4}$ & 4.4, eac$^{4}$ & eac$^{4}$ & 34 \\
ELHC 5 & 1994 Aug$^{6}$ & 8.1 & -11 & $\sim$1.8$^{5}$, eac & filled abs line & 75 \\
ELHC 5 & 2005 Jan 21  & 5.3 & -5.4 & 1.9, eac & weak ebc$^{2}$ & 27 \\
ELHC 6 & 1994 Aug$^{6}$ & 14.7 & -20 & $\sim$2$^{5}$, eac, filled abs line & 75 \\
ELHC 6 & 2005 Jan 21  & 10.7$^{4}$ & -25.6$^{4}$ & 3.3, eac$^{4}$ & eac$^{4}$ & 31 \\
ELHC 7 & 1994 Aug$^{6}$ & 9.8 & -130 & $\sim$7.5, eac & eac & 15 \\
ELHC 7 & 2005 Jan 21 & 9.2 & -106 & 10.8, dpe$^{3}$ & eac & 13 \\
ELHC 8 & 2000 Jan 8$^{7}$ & \nodata & -34 & eac & \nodata & 80 \\
ELHC 8 & 2005 Jan 21  & 6.4 & -35.3 & 6.1, eac & eac & 42 \\
ELHC 11 & 2005 Jan 21  & 8.3$^{4}$ & -17.5$^{4}$ & 3.0, eac$^{4}$ & eac$^{4}$ & 47 \\
ELHC 12 & 2000 Jan 6$^{7}$ & \nodata & \nodata & $\sim$3.1$^{5}$, eac & eac & 30 \\ 
ELHC 12 & 2005 Jan 21 & 11.5 & -34 & 3.8, eac & eac & 36 \\
ELHC 13 & 2000 Mar 9$^{7}$ & \nodata & -35.4 & $\sim$3.1$^{5}$, eac & eac & 70 \\
ELHC 13 & 2005 Jan 21 & 6.6 & -20.2 & 3.8, eac & eac & 43 \\
ELHC 16 & 1998 Jan 6$^{7}$ & \nodata & \nodata & \nodata, eac & \nodata & 70 \\
ELHC 16 & 2005 Jan 21 & 3.0 & -1.9 & 1.9, eac & weak ebc & 35 \\
ELHC 19 & 1998 Jan 6$^{7}$ & \nodata & \nodata & \nodata, eac & \nodata & 40 \\
ELHC 19 & 2005 Jan 21  & 4.8 & -29 & 6.3, eac & eac & 34 \\
ELHC 20 & 1998 Jan 6$^{7}$ & \nodata & \nodata & \nodata, filled abs line & \nodata & 60 \\
ELHC 20 & 2005 Jan 21 & \nodata & 4.5 & weak ebc & abs & 17 \\
ELHC 21 & 1998 Jan 8$^{7}$ & 14.5 & 13.0 & $\sim$1.4$^{5}$, eac &  filled abs & 45 \\
ELHC 21 & 2005 Jan 21  & 13.0 & -8.7 & 1.8, dpe$^{3}$ & weak dpe & 29 \\
ESHC 1 & 1998 Aug 18$^{8}$ & 7.5 & -6.2 & $\sim$1.6$^{5}$, eac & eac & 100 \\
ESHC 1 & 2005 Jan 21  & 3.5 & 0.6 & 1.3, weak eac & abs & 13 \\
ESHC 2 & 1998 Aug 18$^{8}$ & 10.1 & -18.5 & $\sim$2.6$^{5}$, eac & \nodata & 20 \\
ESHC 2 & 2005 Jan 21  & 6.7 & -11.8 & 2.8, eac & weak eac & 16 \\
ESHC 3 & 2005 Jan 21 & 4.0 & -6.8 & 2.4, eac & filled abs line & 24 \\
ESHC 4 & 2005 Jan 21 & 7.8 & -3.9 & 1.5, dpe & weak ebc & 44 \\
ESHC 5 & 2005 Jan 21 & 8.4 & -9.8 & 2.1, dpe & abs & 74 \\
ESHC 6 & 2005 Jan 21 & 9.1 & -33.2 & 4.2, eac & eac & 47 \\
ESHC 7 & 2005 Jan 21  & 5.6 & -26.7 & 5.6, eac & eac & 11 \\
\tableline
\tablecomments{The basic properties of our spectroscopic data are summarized.  Column 5 includes a measure of the ratio of the peak-to-continuum level of the H$\alpha$ emission line along with a description of the line profile.  Column 6 provides a description of the H$\beta$ line profile.  $^{1}$ indicates emission above the continuum, $^{2}$ indicates emission which partially fills in an underlying photospheric absorption line which doesn't emerge above the continuum, $^{3}$ indicates double-peaked emission, $^{4}$ indicates sources which suffer from source confusion, as discussed in Section \ref{confuse}.  $^{5}$ indicates that we estimated the peak-to-continuum ratio from Figures published in literature.  $^{6}$ \citet{lam99} , $^{7}$ \citet{dew05}, $^{8}$ \citet{bea01} }
\end{tabular}
\end{center}
\end{table}

\newpage
\clearpage

\newpage
\clearpage

\newpage
\clearpage


\begin{thebibliography}{}

\bibitem[Ashok et al.(1984)]{ash84} Ashok, N.M., Bhatt, H.C., Kulkarni, P.V., \& Joshi, S.C. 1984, MNRAS, 211, 471
\bibitem[Bary et al.(2009)]{bar09} Bary, J.S., Leisenring, J.M., \& Skrutskie, M.F. 2009, ApJL, 706, 168
\bibitem[Bastien(1988)]{bas88} Bastien, P. 1988, in Polarized Radiation of Circumstellar Origin, ed. 
G.V. Coyne et al. (Vatican City:Vatican Obs), 541
\bibitem[Beaulieu et al.(2001)]{bea01} Beaulieu, J.-P. et al. 2001, A\&A, 380, 168
\bibitem[Bertin \& Arnouts(1996)]{ber96} Bertin, E. \& Arnouts, S. 1996, A\&AS, 117, 393
\bibitem[Bertin et al.(2002)]{ber02} Bertin, Mellier, Y., Radovich, M., Missonnier, G., Didelon, P., \& Morin, B. 2002, in Astronomical Data Analysis Software and Systems XI, ASP Conf Ser (ed D.A. Bohlender, D. Durand, \& T.H. Handley), 281, 228
\bibitem[Bjorkman et al.(2002)]{bjo02} Bjorkman, K.S., Miroshnichenko, A.S., McDavid, D., Pogrosheva, T.M. 2002, ApJ, 573, 812
\bibitem[Bjorkman et al.(2005)]{bjo05} Bjorkman, K.S., Wisniewski, J.P., Bjorkman, J.E., \& Hesselbach, E.N. 2005, Protostars and Planets V, 1286, 8416
\bibitem[Bonanos et al.(2010)]{bon10} Bonanos, A.Z. 2010, AJ, 140, 416
\bibitem[Bonanos et al.(2009)]{bon09} Bonanos, A.Z. 2009, AJ, 138, 1003
\bibitem[Carciofi(2011)]{car11} Carciofi, A.C. 2011, in Active OB Stars: Structure, Evolution, Mass Loss, proc IAU Symp 272 (ed. C Neiner, G Wade, G Meynet, G Peters), in press (astro-ph/1009.3969)
\bibitem[Chiappini et al.(2006)]{chi06} Chiappini, C., Hirschi, R., Meynet, G., Ekstrom, S., Maeder, A., \& Matteucci, F. 2006, A\&A, 449, L27
\bibitem[Clayton et al.(2010)]{cla10} Clayton, G.C. et al. 2010, ApJ, 722, 1131
\bibitem[de Wit et al.(2002)]{dew02} de Wit, W.J., Beaulieu, J.P., \& Lamers, H.J.G.L.M. 2002, A\&A, 395, 829
\bibitem[de Wit et al.(2003)]{dew03} de Wit, W.J., Beaulieu, J.P., Lamers, H.J.G.L.M., Lesquoy, E., \& Marquette, J.B. 2003, A\&A, 410, 199
\bibitem[de Wit et al.(2005)]{dew05} de Wit, W.J., Beaulieu, J.P., Lamers, H.J.G.L.M., Coutures, C., \& Meeus, G. 2005, A\&A, 432, 619
\bibitem[Dougherty \& Taylor(1994)]{dou94} Dougherty, S.M. \& Taylor, A.R. 1994, MNRAS, 269, 1123
\bibitem[Draper et al.(2011)]{dra11} Draper, Z.H. et al. 2011, ApJL, 728, 40
\bibitem[Eiroa et al.(2002)]{eir02} Eiroa, C. et al. 2002, A\&A, 384, 1038
\bibitem[Fischer \& Valenti(2005)]{fis05} Fischer, D.A. \& Valenti, J. 2005, ApJ, 622, 1102
\bibitem[Gandet et al.(2002)]{gan02} Gandet, T.L., Otero, S., Fraser, B., \& West, J.D. 2002, IBVS, 5352
\bibitem[Gonzalez (1997)]{gon97} Gonzalez, G. 1997, MNRAS, 285, 403
\bibitem[Grinin et al.(1991)]{gri91} Grinin, V.P., Kiselev, N.N., Minikulow, N. Kh., Chernova, G.P., \& Voshchinnikov, N.V. 1991, Ap\&SS, 186, 283
\bibitem[Grinin et al.(1994)]{gri94} Grinin, V.P., The, P.S., de Winter, D., Giampapa, M., Rostopchina, A.N., Tambovtseva, L.V., \& van den Ancker, M.E. 1994, 
A\&A, 292, 165
\bibitem[Herbig(1960)]{her60} Herbig, G.H. 1960, ApJS, 4, 337
\bibitem[Herbst \& Shevchenko(1999)]{her99} Herbst, W. \& Shevchenko, V.S. 1999, AJ, 118, 1043
\bibitem[Hillenbrand et al.(1992)]{hil92} Hillenbrand, L.A., Strom, S.E., Vrba, F.J., \& Keene, J. 1992, ApJ, 397, 613
\bibitem[Hubert \& Floquet(1998)]{hub98} Hubert, A.M. \& Floquet, M. 1998, A\&A, 335, 565
\bibitem[Hummel (1998)]{hum98} Hummel, W. 1998, A\&A, 330, 243
\bibitem[Lamers et al.(1999)]{lam99} Lamers, H.J.G.L.M., Beaulieu, J.P., \& de Wit, W.J. 1999, A\&A, 341, 827
\bibitem[McSwain et al.(2009)]{mcs09} McSwain, M.V., Huang, W., \& Gies, D. 2009, ApJ, 700, 1216
\bibitem[Meixner et al.(2006)]{mei06} Meixner, M. et al. 2006, AJ, 132, 2268
\bibitem[Meynet et al.(1994)]{mey94} Meynet, G., Maeder, A., Schaller, G., Schaerer, D., \& Charbonnel, C. 1994, A\&AS, 103, 97
\bibitem[Mink (1997)]{min97} Mink, D.J. 1997, in Astronomical Data Analysis Software and Systems VI, ASP Conf Ser (ed. G Hunt \& H.E. Payne), 125, 249
\bibitem[Miroshnichenko et al.(2003)]{mir03} Miroshnichenko, A.S. et al. 2003, A\&A, 408, 305
\bibitem[Muzerolle et al.(2009)]{muz09} Muzerolle, J. et al. 2009, ApJL, 704, 15
\bibitem[Oudmaijer et al.(2001)]{oud01} Oudmaijer, R.D. et al. 2001, A\&A, 564, 578
\bibitem[Palanque-Delabrouille et al.(1998)]{pal98} Palanque-Delabrouille, N. et al. 1998, A\&A, 332, 1
\bibitem[Persson et al.(1998)]{per98} Persson, S.E., Murphy, D.C., Krzeminski, W., Roth, M., \& Rieke, M.J. 1998, AJ, 116, 2475
\bibitem[Pogodin(1994)]{pog94} Pogodin, M.A. 1994, A\&A, 282, 141
\bibitem[Porter \& Rivinius(2003)]{por03} Porter, J.M. \& Rivinius, T. 2003, PASP, 115, 1153
\bibitem[Reipurth et al.(1996)]{rei96} Reipurth, B., Pedrosa, A., \& Lago, M.T.V.T. 1996, A\&AS, 120, 229
\bibitem[Robitaille et al.(2008)]{rob08} Robitaille, T.P. et al. 2008, AJ, 136, 2413
\bibitem[Skrutskie et al.(2006)]{skr06} Skrutskie, M.F. et al. 2006, AJ, 131, 1163
\bibitem[Stefl et al.(2009)]{ste09} Stefl, S. et al. 2009, A\&A, 504, 929
\bibitem[Urquhart et al.(2011)]{urq10} Urquhart, J.S. et al. 2011, MNRAS, 410, 1237
\bibitem[Waters \& Waelkens(1998)]{wat98} Waters, L.B.F.M. \& Waelkens, C. 1998, ARAA, 36, 233
\bibitem[van der Bliek et al.(2004)]{van04} van der Bliek, N.S. et al. 2004, in Ground-based Istrumentation for Astronomy, Proc SPIE, 5492, 1582
\bibitem[Vieira et al.(1999)]{vie99} Vierira, S.L.A., Pogodin, M.A., \& Franco, G.A.P. 1999, A\&A, 345, 559
\bibitem[Vijh et al.(2009)]{vij09} Vijh, U.P. 2009, AJ, 137, 3139
\bibitem[Underhill \& Doazan(1982)]{und82} Underhill, A. \& Doazan, V. 1982, B Stars with and without Emission Lines (Washington DC:NASA), NASA SP-456
\bibitem[Whitney et al.(2008)]{whi08} Whitney, B.A. et al. 2008, AJ, 136, 18
\bibitem[Wisniewski \& Bjorkman(2006)]{wis06} Wisniewski, J.P. \& Bjorkman, K.S. 2006, ApJ, 652, 458
\bibitem[Wisniewski et al.(2007a)]{wis07} Wisniewski, J.P., Bjorkman, K.S., Magalhaes, A.M., Meade, M.R., \& Pereyra, A. 2007a, ApJ, 671, 2040
\bibitem[Wisniewski et al.(2007b)]{wi07b} Wisniewski, J.P., Kowalski, A.F., Bjorkman, K.S., Bjorkman, J.E., \& Carciofi, A.C. 2007b, ApJL, 656, 21
\bibitem[Wisniewski et al.(2010)]{wis10} Wisniewski, J.P., Draper, Z.H., Bjorkman, K.S., Meade, M.R., Bjorkman, J.E., \& Kowalski, A.F. 2010, ApJ, 709, 1306




\end{thebibliography}
\end{document}